\documentclass[11pt]{article}

\usepackage{geometry} 
\usepackage{graphicx} 

\usepackage{float} 
\usepackage{wrapfig} 
\usepackage{lipsum} 
\usepackage{exscale}
\usepackage{relsize}
\usepackage{indentfirst}
\usepackage{amssymb}
\usepackage{amsfonts}
\usepackage{amsthm}
\usepackage{amssymb}
\usepackage{amsmath}
\usepackage{abstract}
 \usepackage{mathrsfs}
\usepackage{dutchcal}
\usepackage{multirow}
\usepackage{caption}
 \usepackage{hhline}

\graphicspath{{./Pictures/}} 

\numberwithin{equation}{section}

\newtheorem{thm}{Theorem}[section]
\newtheorem{defn}[thm]{Definition}
\newtheorem{lem}[thm]{Lemma}

\newtheorem{prop}[thm]{Proposition}
\newtheorem{remark}[thm]{Remark}
\newtheorem{exam}[thm]{Example}
\newtheorem{conj}[thm]{Conjecture}

\begin{document}

\title{New Constructions  of  Constant Dimension Subspace Codes with Large Sizes\let\thefootnote\relax\footnotetext{
 E-Mail addresses: yunly@mails.ccnu.edu.cn (Yun  Li),  hwliu@ccnu.edu.cn (Hongwei Liu), smesnager@univ-paris8.fr (Sihem Mesnager).}}

\author{Yun Li$^1$,~Hongwei Liu$^2$,~Sihem Mesnager$^3$}
\date{}
\maketitle
\begin{center}
\small $^{1,2}$School of Mathematics and Statistics, Central China Normal University, Wuhan, 430079 China \\
$^{3}$LAGA, Department of Mathematics, Universities of Paris VIII and Paris XIII, CNRS, UMR 7539 and Telecom ParisTech, France
\end{center}
\begin{abstract}

Subspace codes have important applications in random network coding. It is interesting to construct subspace codes with both sizes, and the minimum distances are as large as possible. In particular, cyclic constant dimension subspaces codes have additional properties which can be used to make encoding and decoding more efficient. In this paper, we construct large cyclic constant dimension subspace codes with minimum distances $2k-2$ and $2k$. These codes are contained in $\mathcal{G}_q(n, k)$, where $\mathcal{G}_q(n, k)$ denotes the set of all $k$-dimensional subspaces of $\mathbb{F}_{q^n}$. Consequently, some results in \cite{FW}, \cite{NXG}, and \cite{ZT} are extended.


\end{abstract}

\section[Introduction]{Introduction}

Ahlswede et al.\cite{ACLY} introduced random network coding, which has been proven to be a very effective tool in a non-coherent network. Let $\mathbb{F}_q$ be the finite field with $q$ elements, and let $\mathbb{F}_{q^n}$ be the extension field of degree $n$ over $\mathbb{F}_q$. K\"otter and Kschischang \cite{KK} provided an algebraic approach to random network coding in non-coherent networks, and they considered messages as subspaces of some fixed vector space $\mathbb{F}_q^n$. A code is a set of some subspaces of $\mathbb{F}_q^n$ and codewords are simply subspaces. The authors also gave error correction and the corresponding transmission model in \cite{KK}. Since then, some researches focus on the construction and bounds of subspace codes, e.g. \cite{BEGR}, \cite{BKR}, \cite{GL}, \cite{KKS}, \cite{RT}, \cite{T}, \cite{XF}. Note that the extension field $\mathbb{F}_{q^n}$ over $\mathbb{F}_q$ is also an $\mathbb{F}_q$-vector space with dimension $n$. Researchers are keen to consider messages as subspaces of $\mathbb{F}_{q^n}$ due to that it is algebraically richer than $\mathbb{F}_q^n$, and one can utilize properties of the field extension $\mathbb{F}_{q^n}$. In this paper, we construct some new subspace codes of $\mathbb{F}_{q^n}$ with large sizes.

Let $\mathcal{P}_q (n)$ be the set of all subspaces of $\mathbb{F}_{q^n}$.  Let $\mathcal{G}_q(n,k)$ be the set of all $k$-dimensional subspaces of $\mathbb{F}_{q^n}$ over the finite $\mathbb{F}_q$. For any two subspaces $U$ and $V$ of $\mathcal{P}_q (n)$, the {\it (subspace) distance} between $U$ and $V$ is defined as
\begin{center}
$d(U,V)=\dim U +\dim V -2\dim(U\cap V)$,
\end{center}
where $U\cap V$ denotes the intersection subspace of $U$ and $V$. The subspace distance is actually a metric of $\mathcal{P}_q (n)$, and we call  $\mathcal{P}_q (n)$  a metric space with this distance.
 A non-empty subset $\mathcal{C}\subseteq \mathcal{P}_q(n)$ with this metric is called a  {\it subspace code}. Moreover, if  $\mathcal{C}\subseteq \mathcal{G}_q(n,k)$, then $\mathcal{C}$ is called a {\it constant dimension ($k$-dimensional) subspace code}.  The {\it minimum (subspace) distance} of a subspace code $\mathcal{C}$ is defined by
 $$d(\mathcal{C})=\min\{d(U,V)\,|\, U,V \in\mathcal{C}, U\neq V\}.$$

The set $\textup{orb}(U)=\{\alpha U\,|\,\alpha\in \mathbb{F}_{q^n}^*\}$ is called the {\it orbit} of $U$. For any nonzero $\alpha$ in $\mathbb{F}_{q^n}$, we call $\alpha U$ the {\it cyclic shift} of $U$. These orbits are subspace codes that arise as an orbit of a subspace in $\mathbb{F}_{q^n}$ under the finite group $\mathbb{F}_{q^n}^*$. When the subspace code $\mathcal{C}$ is a union of some orbits, then we call $\mathcal{C}$ {\it cyclic}. It is known that a cyclic single orbit code $\textup{orb}(U)$ has minimum distance $2k$ if and only if $U$ is a cyclic shift of $\mathbb{F}_{q^k}$, and $k$ must be a divisor of $n$ (\cite{OO}). The size of this single orbit code is $\frac{q^n-1}{q^k-1}$. In other cases, the minimum distance of $\textup{orb}(U)$ is less than or equal to $2k-2$, and the size of $\textup{orb}(U)$ is less than or equal to $\frac{q^n-1}{q-1}$. Therefore, many researchers focus on constructing cyclic $k$-dimensional (constant dimension) subspace code, a union of as many single simple orbit codes whose minimum distances can reach $2k-2$. Furthermore, the size of this cyclic constant dimension subspace code is equal to $m \frac{q^n-1}{q-1}$, where $m$ is a positive integer.

A {\it linearized polynomial} ($q$-polynomial) $f(x)$ is a polynomial with the form $f(x)=\sum\limits_{i=0}^k a_i x^{q^i}\in\mathbb{F}_{q^n} [x]$ and $k$ is called the {\it $q$-degree} of $f(x)$ if $a_k \neq 0$. A {\it subspace polynomial} $f(x)$ is a monic linearized polynomial with all roots are single, and $f(x)$ splits in $\mathbb{F}_{q^n}$ \cite{LN}. All roots of a subspace polynomial $f(x)$ form a linear $\mathbb{F}_q$ subspace $U$ of $\mathbb{F}_{q^n}$. For an arbitrary subspace $U\in\mathcal{G}_q (n,k)$, $f(x)=\prod\limits_{u\in U} (x-u)$ is a subspace polynomial over $\mathbb{F}_{q^n}$. Thus there is a one-to-one correspondence between subspaces in $\mathcal{G}_q (n, k)$ and subspace polynomials with $q$-degree $k$.
Ben-Sasson et al. constructed a cyclic constant dimension subspace code with size $\frac{q^n-1}{q-1}$ and minimum distance $2k-2$ by using some subspace polynomials in \cite{BEGR}.
 In \cite{OO}, Otal and \"Ozbudak constructed a cyclic constant dimension subspace code which is a union of $r$ distinct cyclic single orbit codes that constructed by $r$ distinct subspace polynomials, and thus this cyclic constant dimension code has size $r\frac{q^n-1}{q-1}$ and minimum distances $2k-2$. Chen and Liu constructed several cyclic constant subspace codes in \cite{CL} by using more general subspace polynomials. In general, codes constructed by using subspace polynomials are contained in $\mathcal{G}_q(n, k)$ such that $k$ is much less than $n$.

Roth et al. \cite{RRT} provided new methods to construct subspace codes using Sidon spaces. They have solved some cases of Conjecture \ref{conj2.4} \cite{TMBR} for $n>2k$. Niu et al. \cite{NYW} gave a new construction that yields cyclic subspace codes that have more codewords than previous constructions with the tool of linearized polynomials. Feng and Wang \cite{FW} constructed some large subspace codes in $\mathcal{G}_q(3k, k)$ and $\mathcal{G}_q(7k, 2k)$. Recently, Li and Liu \cite{LL} constructed several new subspace codes by using the sum of some Sidon spaces. Our work in this paper aims to extend some previous work in \cite{FW, NXG, ZT}, and obtain some new constant dimension subspace codes with large sizes. Previously known results about the sizes of constant dimension subspace codes, together with our main results in this paper, are collected in Table 1.
\begin{table}[h]
\centering
\small
\caption*{\textbf{Table 1} Sizes of constant dimension subspace codes with distances $2k-2$ and $2k$}
\vskip 2mm

\begin{tabular}{c|c|c|c}

\hline
 & Sizes  & Distances & References\\
\hline
\multirow{3}*{$\frac{n}{k}>2$} & $e q^k \frac{q^n-1}{q-1} (e=\lceil \frac{n}{2k}\rceil -1)$ &\multirow{3}*{$2k-2$} & \cite{FW} \\
\cline{2-2}\cline{4-4}
 ~                                        & $2(q^k-1) \frac{q^n-1}{q-1}+\frac{q^n-1}{q^k-1}$ & ~ & \cite{NXG} \\
\cline{2-2}\cline{4-4}
 ~                                        & $e q^k (q^n-1)+\frac{q^n-1}{q^k-1} (e=\lceil \frac{n}{2k}\rceil-1)$ & ~ & {\bf Theorem 3.4} (New) \\
\hline
\multirow{4}*{$\frac{n}{k}>4$} & $leq^k\frac{q^n-1}{q-1}  (e=\lceil\frac{n}{4k}\rceil -1, l\leq k)$ & \multirow{2}*{$2k-2$} & \cite{NXG} \\
\cline{2-2}\cline{4-4}
~                                         & $q^{2k} \frac{q^n-1}{q-1} $ & ~ & {\bf Theorem 3.7} (New) \\
\cline{2-4}
~                                         & $q^n-1$ & \multirow{2}*{$2k$} & \cite{ZT} \\
\cline{2-2}\cline{4-4}
~                                         & $e q^k \frac{q^n-1}{q-1} (e=\lceil \frac{n}{4k}\rceil -1)$ & ~ & {\bf Theorem 3.11} (New) \\
\hline
\end{tabular}
\end{table}

This paper is organized as follows. In Section 2, we present some preliminary results. In Section 3, we provide a construction of a cyclic subspace code (see Theorem \ref{thm3.4}) with minimum distance $2k-2$ and size $e q^k (q^n-1)+\frac{q^n-1}{q^k-1}$, where $n$ is a multiple of $k$ with $n >2k$. And we construct a cyclic subspace code (see Theorem \ref{thm3.7}) with minimum distance $2k-2$ and size $q^{2k} \frac{q^n-1}{q-1}$, as the case $n>4k$.
We emphasize that this size is much larger than the sizes of codes constructed in \cite{NXG}. We also obtain a cyclic subspace code (see Theorem \ref{thm3.11}) with minimum distance $2k$ and size $e q^k\frac{q^n-1}{q-1}$ which is much larger than sizes of codes constructed in \cite{ZT}.

\section[Preliminaries]{Preliminaries}

In this section, we give some notions and propositions which will be used in the following section.

Let $\mathbb{F}_q$ be the finite field of order $q$, where $q$ is a prime power. Let $\mathbb{F}_{q^n}$ be the extension field of $\mathbb{F}_q$. Then $\mathbb{F}_{q^n}$ can be viewed as an $n$-dimensional $\mathbb{F}_q$-vector space. Let $\mathbb{F}^n_q=\{(a_1,\cdots, a_n)\,|\, a_i\in \mathbb{F}_q\}$ be the $n$-dimensional vector space over $\mathbb{F}_q$. Let $U, V$ be two subspace of $\mathbb{F}_{q^n}$, the {\it sum} and {\it product} of $U$ and $V$ are defined as follows, respectively
$$
U+V=\{u+v\,|\, u\in U, v\in V\}, \quad UV=\langle\{uv\,|\, u\in V, v\in V\} \rangle,
$$
where $\langle S\rangle$ denote the subspace generated by the subset $S$ of $\mathbb{F}_{q^n}$. We denote the direct sum of $U$ and $V$ by $U\oplus V$. For a set $S$, the cardinality (size) of $S$ is denoted by $|S|$. For a real number $r$, we let $\lceil r\rceil$ denote the least integer greater than or equal to $r$.

The cardinality of a cyclic simple orbit subspace code can be determined by the following proposition.

\begin{prop}\label{prop2.1}
\cite{OO}
Let $U\in \mathcal{G}_q (n, k)$. Then $\mathbb{F}_{q^d}$ is the largest field such that $U$ is also $\mathbb{F}_{q^d}$-linear if and only if
$$\mid\textup{orb}(U)\mid=\frac{q^n-1}{q^d-1}.$$
\end{prop}

In the following, it has been shown that the construction of cyclic simple orbit codes with size $\frac{q^n-1}{q-1}$ and minimum distance $d(\mathcal{C}) = 2k-2$ can be turned into the construction of Sidon spaces.

\begin{defn}\label{def2.2}
\cite{BSZ}
A subspace $U\in\mathcal{G}_q(n,k)$ is called a Sidon space if for any nonzero elements $a,b,c,d \in U$,
  if $ab=cd$, then $\{a \mathbb{F}_q, b \mathbb{F}_q\}=\{c \mathbb{F}_q, d \mathbb{F}_q\}$.
 \end{defn}

\begin{prop}\label{prop2.3}
\cite{RRT}
 For a subspace $U \in \mathcal{G}_q(n,k)$, the code $\textup{orb}(U)$ is of size $\frac{q^n-1}{q-1}$
 and minimum distance $2k-2$ if and only if $U$ is a Sidon space.
\end{prop}

\begin{conj}\label{conj2.4}
\cite{TMBR}
For any prime power $q$ and positive integers $k$ and $n>2k$, there exists a cyclic subspace code (cyclic single orbit code) $\mathcal{C}\subseteq \mathcal{G}_q (n, k)$ with size $\mid \mathcal{C}\mid=\frac{q^n-1}{q-1}$
and minimum distance $d(\mathcal{C})=2k-2.$
\end{conj}

Roth et al. \cite{RRT} proved the conjecture for $q \geq 3$ in the case where $n$ is even and $n\geq 2k$, and for $q\geq 2$ in the case where $n>2k$. \cite{LL} extended some results in \cite{RRT}.

\begin{prop}\label{prop2.5}
\cite{RRT}
The following two conditions are equivalent for any distinct subspaces $U$ and $V$ in $\mathcal{G}_q(n,k)$.\\
 \indent (1) $\dim(U\cap \alpha V)\leq 1,$ for any $\alpha\in \mathbb{F}_{q^n}^*.$\\
 \indent (2) For any nonzero $a,c\in U $ and nonzero $b,d\in V$, the equality $ab=cd$ implies that $a \mathbb{F}_q=c \mathbb{F}_q$
 and $b \mathbb{F}_q=d \mathbb{F}_q$.
\end{prop}

 The above proposition can be used to prove that $\dim(U\cap \alpha V)\leq 1,$ for any $\alpha\in \mathbb{F}_{q^n}^*$, for some distinct subspaces $U$ and $V$ of $\mathbb{F}_{q^n}$. The following proposition shows that the sum of two distinct Sidon subspaces $U$ and $V$ of $\mathbb{F}_{q^n}$ that satisfy some conditions is again a Sidon space. These two propositions are used frequently in Section 3.

\begin{prop}\label{prop2.6}
\cite{LL}
Let $U,V\leq\mathbb{F}_{q^n}$ be two Sidon spaces. If $U$ and $V$ satisfy the following two conditions,

\indent (1)~ $(U+V)^2=U^2\oplus UV\oplus V^2$,\\
\indent (2)~ $\dim(U\cap \alpha V)\leq 1$, for any $\alpha\in \mathbb{F}_{q^n}^*,$\\
  then $U+V$ is also a Sidon space.
 \end{prop}

\section[Main results]{Construction of several large subspace codes}

In this section, we construct several large cyclic constant dimension subspace codes. Before starting it, we recall some basic definitions and facts.

Let $\xi$ be a primitive element of $\mathbb{F}_{q^k}$.
Let $G$ be the quotient group $\mathbb{F}_{q^k}^*/\langle\xi^{q-1}\rangle$, then $G$ is a cyclic group. Assume the generator of $G$ is $\overline{\xi}$. Then
$$G=\mathbb{F}_{q^k}^*/\langle\xi^{q-1}\rangle=\langle\overline{\xi}\rangle=\{\overline{\xi}^i\mid 0\leq i\leq q-2, i\in\mathbb{Z}^+\}$$ is a finite cyclic group with cardinality $q-1$.

For any positive integer $k>1$ and $n=rk$ with $r>2$  we let $e=\lceil  r/2 \rceil -1$.  Suppose $\gamma$ is a root of an irreducible polynomial of degree $r$ over $\mathbb{F}_{q^k}$. Let
$$U=\{u+(\tau u^q+u)\delta \gamma^{l}\mid u\in\mathbb{F}_{q^k}\},$$
where $\tau\in G$, $\delta\in\mathbb{F}_{q^k}^*$, $1\leq l\leq e$. It is easy to calculate that there are $e(q-1)(q^k-1)$ subspaces have the same form with $U$. Then we can arrange these subspaces in order by
$U_1,U_2,\cdots, U_{e(q-1)(q^k-1)}.$ Let
$$U_i=\{u+(\tau_i u^q+u)\delta_i \gamma^{l_i}\mid u\in\mathbb{F}_{q^k}\},$$
where $\tau_i\in G$, $\delta_i\in\mathbb{F}_{q^k}^*$, $1\leq l_i\leq e$, for any $1\leq i\leq e(q-1)(q^k-1)$.

\begin{lem}\label{lem3.1}

Keep the notation above. Let $k\geq 2$ be an integer, $n=rk, r>2\in\mathbb{Z}^{+}$, and $e=\lceil r/2 \rceil -1$. Let
$\mathcal{C}_i=\{\alpha U_i\mid \alpha\in\mathbb{F}_{q^n}^*\}.$ Then the subspace code
 $$\mathcal{C}=\bigcup\limits_{i=1}^{e(q-1)(q^k-1)} \mathcal{C}_i$$
 is a cyclic constant dimension subspace code with cardinality $e (q^k-1) (q^n-1)$ and minimum distance $2k-2$.

\begin{proof}
Firstly, we prove that the minimum distance of $\mathcal{C}$ is equal to $2k-2$. By Proposition \ref{prop2.5}, we just need to prove that $\dim(U_i\cap \alpha U_j)\leq 1$, for any $\alpha\in\mathbb{F}_{q^n}^*,$ and for any positive integers $1\leq i, j\leq e(q-1)q^k$. Let
\begin{center}
$\mathcal{u}=u+(\tau_i u^q+u)\delta_i\gamma^{l_i}, \,\, \mathcal{u}'=u'+(\tau_i u'^q+ u')\delta_i\gamma^{l_i}\in U_i,$
\end{center}
and
\begin{center}
 $\mathcal{v}=v+(\tau_j v^q+ v)\delta_j\gamma^{l_j}, \,\, \mathcal{v}'=v'+(\tau_j v'^q+v')\delta_j \gamma^{l_j}\in U_j$
 \end{center}
be four nonzero elements such that $\mathcal{u}\mathcal{v}=\mathcal{u}'\mathcal{v}'$. Recall that $\gamma$ is a root of an irreducible polynomial of degree $r$ over $\mathbb{F}_{q^k}$, and $n/k=r$. We know that the elements in the set $\{\gamma^{i}\mid 0\leq i\leq r-1\}$ are linear independent over $\mathbb{F}_{q^k}$.

If $l_i\neq l_j,$ the equation $\mathcal{u}\mathcal{v}=\mathcal{u}'\mathcal{v}'$ leads to
\begin{center}
$uv=u'v',$\\
$\tau_i \delta_i u^q v=\tau_i\delta_i u'^q v'$.
\end{center}
Thus, we have $(\frac{u}{u'})^q=(\frac{u^q}{u'^q})=\frac{v'}{v}=\frac{u}{u'}$. This implies that $\frac{u}{u'}=\lambda\in\mathbb{F}_q^*.$ This means that
$\frac{\mathcal{v}'}{\mathcal{v}}=\frac{\mathcal{u}}{\mathcal{u}'}=\lambda\in\mathbb{F}_q^*.$

If $l_i= l_j,$ the equation $\mathcal{u}\mathcal{v}=\mathcal{u}'\mathcal{v}'$ leads to\begin{equation}\label{eq3.1}
uv=u'v',
\end{equation}
\begin{equation}\label{eq3.2}
\tau_j \delta_j u v^q+\tau_i \delta_i u^q v=\tau_j \delta_j u' v'^q+\tau_i\delta_i u'^q v',
\end{equation}
\begin{equation}\label{eq3.3}
\tau_j\delta_j\delta_i u v^q+\tau_i \delta_i\delta_j u^q v=\tau_j\delta_j\delta_i u' v'^q+\tau_i \delta_i\delta_j u'^q v'.
\end{equation}

{\bf Case 1}. $\delta_i\neq\delta_j.$ By Equations \ref{eq3.2} and \ref{eq3.3}, we have $u^q v=u'^q v'$. Thus, we know that $\frac{v'}{v}=\frac{u^q}{u'^q}=\frac{u}{u'}=\lambda\in\mathbb{F}_q^*.$
This means that $\frac{\mathcal{u}}{\mathcal{u'}}=\frac{\mathcal{v'}}{\mathcal{v}}=\lambda\in\mathbb{F}_q^*.$

{\bf Case 2}. $\delta_i=\delta_j.$ Let
\begin{center}
$f(x)=(u+(\tau_i u^q+u)\delta_i  x)(v+(\tau_j v^q+v)\delta_j  x)$
\end{center}
and
\begin{center}
$f'(x)=(u'+(\tau_i u'^q+u')\delta_i  x)(v'+(\tau_j v'^q+v') \delta_j x)$
\end{center}
be two polynomials in $\mathbb{F}_{q^k}[x]$. Since $1, \gamma,\cdots,\gamma^{r-1}$ are linear independent over $\mathbb{F}_{q^k}$ and $1\leq l_i=l_j\leq e$, then $\mathcal{u}\mathcal{v}=\mathcal{u}'\mathcal{v}'$ implies $f(x)=f'(x)$, which leads to that $f(x)$ and $f'(x)$ have same roots in some extension field of $\mathbb{F}_{q^k}$.

{\bf Subcase 2.1}. $\delta_i=\delta_j, \tau_i\neq \tau_j.$ Since $\tau_i\neq \tau_j$, we claim  that $\tau_i u^{q-1}+1$ and $\tau_j v^{q-1}+1$ can not both be zeros. In fact, if $\tau_i u^{q-1}+1=0=\tau_j v^{q-1}+1$, we can obtain that  $\tau_i u^{q-1}=\tau_j v^{q-1}$, which means that $\frac{\tau_i}{\tau_j}=(\frac{v}{u})^{q-1}\in\langle\xi^{q-1}\rangle$. Since $\tau_i\neq\tau_j\in G=\mathbb{F}_{q^k}^*/\langle\xi^{q-1}\rangle$, then we know that this is a contradiction.

{\bf (i)} Assume that both $\tau_i u^{q-1}+1$ and $\tau_j v^{q-1}+1$ are nonzeros, then $f(x)$ has roots $\frac{1}{(\tau_i u^{q-1}+1)\delta_i}$ and $\frac{1}{(\tau_i v^{q-1}+1)\delta_i}$, and $f'(x)$ has roots $\frac{1}{(\tau_i u'^{q-1}+1)\delta_i}$ and $\frac{1}{(\tau_i v'^{q-1}+1)\delta_i}$.

{\bf (i.a)}. If $\frac{1}{(\tau_i u^{q-1}+1)\delta_i}=\frac{1}{(\tau_i u'^{q-1}+1)\delta_i}$, then we have $u^{q-1}=u'^{q-1}$, which leads to $\frac{u}{u'}=\lambda\in\mathbb{F}_q^*.$
We obtain that $\frac{\mathcal{u}}{\mathcal{u'}}=\frac{\mathcal{v'}}{\mathcal{v}}=\lambda\in\mathbb{F}_q^*.$

{\bf (i.b)}. If $\frac{1}{(\tau_i u^{q-1}+1)\delta_i}=\frac{1}{(\tau_j v'^{q-1}+1)\delta_j}$, then we have $\tau_i u^{q-1}=\tau_j v'^{q-1}$, which leads to $\frac{\tau_i}{\tau_j}=(\frac{v'}{u})^{q-1}\in\langle\xi^{q-1}\rangle.$ Since $\tau_i\neq \tau_j\in G,$ we know that $\frac{\tau_i}{\tau_j}\notin\langle\xi^{q-1}\rangle.$ This is a contradiction.

{\bf (ii)} Assume that there is exactly one of these two elements $\tau_i u^{q-1}+1$ and $\tau_j v^{q-1}+1$ is zero. Without loss of generality, we can suppose that $\tau_i u^{q-1}+1=0$ and $\tau_j v^{q-1}+1\neq 0$, then it is easy to know that $\tau_i u'^{q-1}+1=0$ and $\tau_j v'^{q-1}+1\neq 0$. Thus we have $\frac{u}{u'}\in\mathbb{F}_q^*$, which means that $\frac{\mathcal{u}}{\mathcal{u'}}=\frac{\mathcal{v'}}{\mathcal{v}}=\lambda\in\mathbb{F}_q^*.$ The other case is similar.

{\bf Subcase 2.2}. $\delta_i=\delta_j, \tau_i=\tau_j.$

{\bf (i)}  Assume that both $\tau_i u^{q-1}+1$ and $\tau_i v^{q-1}+1$ are nonzeros.

{\bf (i.a)}. If $\frac{1}{(\tau_i u^{q-1}+1)\delta_i}=\frac{1}{(\tau_i u'^{q-1}+1)\delta_i}$, then we have $u^{q-1}=u'^{q-1}$, which leads to $\frac{u}{u'}=\frac{\mathcal{u}}{\mathcal{u}'}=\frac{\mathcal{v}'}{\mathcal{v}}=\lambda\in\mathbb{F}_q^*.$

{\bf (i.b)}. If $\frac{1}{(\tau_i u^{q-1}+1)\delta_i}=\frac{1}{(\tau_i v'^{q-1}+1)\delta_i}$, then we have $u^{q-1}=v'^{q-1}$, which leads to $\frac{u}{v'}=\frac{\mathcal{u}}{\mathcal{v}'}=\frac{\mathcal{u}'}{\mathcal{v}}\in\mathbb{F}_q^*.$

{\bf (ii)} Assume that there is exactly one of these two elements $\tau_i u^{q-1}+1$ and $\tau_i v^{q-1}+1$ is zero. Then there is exactly one of $\tau_i u'^{q-1}+1$ and $\tau_i v'^{q-1}+1$ is zero. We obtain that $\frac{u}{u'}\in\mathbb{F}_q^*$ or $\frac{u}{v'}\in\mathbb{F}_q^*$.

{\bf (iii)} Assume that both $\tau_i u^{q-1}+1$ and $\tau_i v^{q-1}+1$ are zeros. Then we have $\tau_i u^{q-1}+1=\tau_i v^{q-1}+1=\tau_i u'^{q-1}+1=\tau_i v'^{q-1}+1$, thus we have $u^{q-1}=v^{q-1}=u'^{q-1}=v'^{q-1}.$ Thus $\frac{u}{u'}\in\mathbb{F}_q^*.$

Therefore, by Definition \ref{def2.2} and Proposition \ref {prop2.5}, we conclude that $\dim(U_i\cap \alpha U_j)\leq 1$, for any $\alpha\in\mathbb{F}_{q^n}^*$ and for any $1\leq i, j\leq e(q-1)(q^k-1)$.
Hence the minimum distance of $\mathcal{C}$ equals to $2k-2$ and the cardinality of $\mathcal{C}$ is equal to $e (q-1) (q^k-1)\frac{q^n-1}{q-1}=e (q^k-1)(q^n-1).$
\end{proof}
\end{lem}

\begin{remark}
From Lemma~\ref{lem3.1}, we know that $U_i$, for any $1\leq i\leq e(q-1)(q^k-1)$, are all Sidon spaces.
\end{remark}

\begin{lem}\label{lem3.3}
Let $k\geq 2$ be an integer, and $n=rk, r>2\in\mathbb{Z}^{+}$, $e=\lceil r/2 \rceil -1$. Let
$$V_i=\{u+\eta_i u^q \gamma^{l_i}\mid u\in\mathbb{F}_{q^k}\},\,\, i=1,\cdots, e(q-1),$$
where $\eta_i\in G$ and $1\leq l_i\leq e$. Then the code
$$\mathcal{C}=\bigcup\limits_{i=1}^{q-1} \{\alpha V_i\mid \alpha\in\mathbb{F}_{q^n}^*\}$$
is a cyclic constant dimension subspace code with cardinality $e(q^n-1)$ and minimum distance $2k-2$.
\begin{proof}
Since  $r>2$, we know that $1,\gamma,\cdots,\gamma^{r-1}$ are linear independent over $\mathbb{F}_{q^k}$. Let
$$\mathcal{u}=u+\eta_i u^q\gamma^{l_i}, \mathcal{u}'=u'+\eta_i u'^q\gamma^{l_i}\in V_i, \mathcal{v}=v+\eta_j v^q\gamma^{l_j}, \mathcal{v'}=v'+\eta_j v'^q\gamma^{l_j}\in V_j$$ be four nonzero elements such that $\mathcal{u}\mathcal{v}=\mathcal{u'}\mathcal{v'}.$ Firstly, we know that $uv=u'v'$.

If $l_i\neq l_j$, we have $\eta_i u^q v=\eta_i u'^q v'$, thus we obtain that $$\frac{v}{v'}=\frac{u'}{u}=\frac{\mathcal{u'}}{\mathcal{u}}=\frac{\mathcal{v}}{\mathcal{v}'}=\lambda\in\mathbb{F}_q^*.$$

If $l_i= l_j$, then $\mathcal{u}\mathcal{v}=\mathcal{u}'\mathcal{v}'$ implies that
\begin{center}
$f(x)=(u+\eta_i u^q x)(v+\eta_j v^q x)=(u'+\eta_i u'^q x)(v'+\eta_j v'^q x)=f'(x)\in\mathbb{F}_{q^k}[x],$
\end{center}
and $f(x)$ and $f'(x)$ have same roots.

{\bf Case 1}. $\eta_i\neq\eta_j.$
If the root $\frac{1}{\eta_i u^{q-1}}$ of $f(x)$ equals to the root $\frac{1}{\eta_i u'^{q-1}}$, then we have $\frac{u}{u'}\in\mathbb{F}_q^*$. If the root $\frac{1}{\eta_i u^{q-1}}$ of $f(x)$ equals to the root $\frac{1}{\eta_j v'^{q-1}}$, then we have
$\frac{\eta_i}{\eta_j}=(\frac{v'}{u})^{q-1}\in\mathbb{F}_q^*$, and thus $\eta_i=\eta_j$. This is a contradiction.

{\bf Case 2}. $\eta_i=\eta_j.$
It is easy to prove that $V_i, i=1,\cdots, e(q-1)$ are all Sidon spaces.

From Proposition \ref{prop2.5}, we conclude that $\dim(V_i\cap \alpha V_j)\leq 1$, for any $\alpha\in\mathbb{F}_{q^n}^*$ and for any $1\leq i, j\leq e(q-1)$.
Therefore,  the minimum distance of $\mathcal{C}$ equals to $2k-2$ and the cardinality of $\mathcal{C}$ is equal to $e (q-1) \frac{q^n-1}{q-1}=e(q^n-1).$
\end{proof}
\end{lem}

Combine the results in Lemmas \ref{lem3.1} and \ref{lem3.3}, we have the following theorem.
\begin{thm}\label{thm3.4}
Let $\mathcal{C}_1$ and $\mathcal{C}_2$ be the codes constructed in Lemmas \ref{lem3.1} and \ref{lem3.3}, respectively. And let $\mathcal{C}_3=\textup{orb}(\mathbb{F}_{q^k})$$=\{\alpha \mathbb{F}_{q^k}\mid \alpha\in\mathbb{F}_{q^n}^*\}$. Then $\mathcal{C}=\mathcal{C}_1\cup\mathcal{C}_2\cup\mathcal{C}_3$ is a cyclic constant dimension subspace code with cardinality
$$eq^k(q^n-1)+\frac{q^n-1}{q^k-1}$$ and minimum distance $2k-2$.

\begin{proof}
Firstly, it is easy to verify that $\dim(U_i\cap\alpha \mathbb{F}_{q^k})\leq 1$ and $\dim(V_j\cap\alpha\mathbb{F}_{q^k})\leq 1$ for any $U_i$ given in Lemma \ref{lem3.1} and $V_i$ given in Lemma \ref{lem3.3}. We only need to prove that $\dim(U_i\cap\alpha V_j)\leq 1$ for any $i=1\cdots e(q-1)(q^k-1), j=1,\cdots, e(q-1), \alpha\in\mathbb{F}_{q^n}^*$.  Let
$$\mathcal{u}=u+(\tau_i u^q+u)\delta_i\gamma^{l_i}, \mathcal{u'}=u'+(\tau_i u'^q+u')\delta_i\gamma^{l_i}\in U_i, \mathcal{v}=v+\eta_j v^q\gamma_{l_j},\mathcal{v'}=v'+\eta_j v'^q\gamma_{l_j}\in V_j$$ be four nonzero elements such that $\mathcal{u}\mathcal{v}=\mathcal{u'}\mathcal{v}'$. It is easy to obtain that
$$uv=u'v',$$
$$\delta_i\eta_j uv^q=\delta_i\eta_j u'v'^q.$$
Thus, we have $\frac{u}{u'}=\frac{v'}{v}=(\frac{v'}{v})^{q-1}=\lambda\in\mathbb{F}_q^*.$ By Proposition \ref{prop2.5}, we conclude that $\dim(U_i\cap\alpha V_j)\leq 1$, for any $1\leq i\leq e(q-1)(q^k-1) ,1\leq j\leq e(q-1).$
\end{proof}
\end{thm}

\begin{exam}
Let $n=3\times5=15$, $q=2^2$. The biggest divisor $k$ satisfies that $n/k>2$ is equal to $5$. Thus $n/k=3>2,$ and $e=\lceil 3/2\rceil -1=1$.  The code constructed in Theorem 3.1 of \cite{FW} has size
$$eq^k\frac{q^n-1}{q-1}=2^{10}\frac{2^{30}-1}{2^2-1}.$$
The code constructed in Theorem 3.8 of \cite{NXG} has size
$$2(q^k-1) \frac{q^n-1}{q-1}+\frac{q^n-1}{q^k-1}=2(2^{10}-1) \frac{2^{30}-1}{2^2-1}+\frac{2^{30}-1}{2^{10}-1}>2^{10}\frac{2^{30}-1}{2^2-1}.$$
By Theorem \ref{thm3.7} in this paper, we can construct a cyclic subspace code with size
$$e q^k (q^n-1)+\frac{q^n-1}{q^k-1}=3\times 2^{10}\frac{2^{30}-1}{2^2-1}+\frac{2^{30}-1}{2^{10}-1}>2(2^{10}-1) \frac{2^{30}-1}{2^2-1}+\frac{2^{30}-1}{2^{10}-1}.$$

 \end{exam}

In some special cases, the biggest divisor $k$ of $n$ satisfies $n/k>2$ and $n/k>4$. In this case, we can construct cyclic constant dimension subspace codes with size $q^{2k}\frac{q^n-1}{q-1}$ in the following part of this section.
\begin{lem}\label{lem3.6}
Let $k\geq 2$ be a positive integer, $n=rk, r>4\in\mathbb{Z}^{+}$, and let
\begin{center}
$U_i=\{u+(u^q+u)\tau_i \gamma+(u^q+ u)\delta_i\gamma^{2}\mid u\in\mathbb{F}_{q^k}\}, i=1,\cdots,(q^k-1)^{2}$,
\end{center}
where $\tau_i$, $\delta_i\in\mathbb{F}_{q^k}^*$. Then the code
 \begin{center}
 $\mathcal{C}=\bigcup\limits_{i=1}^{(q^k-1)^2} \mathcal{C}_i, \,\,\mbox{where}\,\,\mathcal{C}_i=\{\alpha U_i\mid \alpha\in\mathbb{F}_{q^n}^*\}, 1\le i\le (q^k-1)^2, $
 \end{center}
 is a cyclic constant dimension subspace code with cardinality $(q^k-1)^2 \frac{q^n-1}{q-1}$ and minimum distance $2k-2$.
\begin{proof}
Let
$$\mathcal{u}=u+(u^q+ u)\tau_i\gamma+(u^q+ u)\delta_i\gamma^{2}, \mathcal{u}'=u'+(u'^q+ u')\tau_i\gamma+(u'^q+u')\delta_i \gamma^{2}\in U_i,$$
$$\mathcal{v}=v+(v^q+ v)\tau_j\gamma+(v^q+v)\delta_j \gamma^{2}, \mathcal{v}'=v'+(v'^q+ v')\tau_j\gamma+(v'^q+v')\delta_j \gamma^{2}\in U_j$$
 be four nonzero elements such that $\mathcal{u}\mathcal{v}=\mathcal{u}'\mathcal{v}'$. Firstly,  it is easy to obtain that
\begin{center}
$uv=u'v'$,\\
$\tau_i u^q v+\tau_j uv^q=\tau_i u'^q v'+\tau_j u'v'^q$,\\
$\delta_i u^q v+\delta_j u v^q=\delta_i u'^q v'+\delta_j u' v'^q,$\\
$u^q v +u v^q=u'^q v' +u' v'^q.$
\end{center}

{\bf Case 1}.  $\tau_j\neq\tau_i$.
It is easy to obtain that $(\tau_j-\tau_i) u v^q=(\tau_j-\tau_i)u'v'^q.$ Thus we have $\frac{u}{u'}=\frac{v'}{v}=(\frac{v'}{v})^q=\lambda\in\mathbb{F}_q^*.$

{\bf Case 2}.  $\tau_j=\tau_i$ and $\delta_i\neq \delta_j.$
It is easy to obtain that $(\delta_j-\delta_i) u v^q=(\delta_j-\delta_i)u'v'^q.$ This case is similar to the case when $\tau_j\neq\tau_i.$

{\bf Case 3}.  $\tau_i= \tau_j$ and $\delta_j=\delta_i.$
Assume that $\frac{u}{u'}=\frac{v'}{v}=\lambda\neq 0.$ Then we have $\lambda u' v^q+\lambda^q u'^q v=\lambda^q u' v^q+\lambda u'^q v.$ This means that $(\lambda-\lambda^q)u' v^q=(\lambda-\lambda^q)u'^q v.$

If $\lambda-\lambda^q=0, $ then $\frac{\mathcal{u}}{\mathcal{u'}}=\frac{u}{u'}=\lambda\in\mathbb{F}_q^*.$ If $\lambda-\lambda^q\neq 0$, then $\frac{u'}{v}=(\frac{u'}{v})^q\in\mathbb{F}_q^*.$

By Definition \ref{def2.2} and Proposition \ref{prop2.5}, we conclude that $\mathcal{C}$ is a cyclic constant dimension subspace code with cardinality $(q^k-1)^2 \frac{q^n-1}{q-1}$ and minimum distance $2k-2$.
\end{proof}
\end{lem}

\begin{thm}\label{thm3.7}
Let
 $$V_j=\{u+u^q\tau'_j \gamma+(u^q+ u)\delta'_j\gamma^{2}\mid u\in\mathbb{F}_{q^k}\}, j=1,\cdots,q^k-1$$
  and
  $$W_l=\{v+(v^q+v)\tau''_l \gamma+v^q\delta''_l\gamma^{2}\mid v\in\mathbb{F}_{q^k}\}, l=1,\cdots,q^k-1$$
  where $\tau'_j, \tau''_l, \delta'_j, \delta''_l\in\mathbb{F}_{q^k}^*$,
   and
   $$U=u+u^q\gamma+u^q\gamma^2.$$
We denote the code constructed in Lemma \ref{lem3.6} by $\mathcal{C}_1$.  Let
$$\mathcal{C}_2=\bigcup\limits_{j=1}^{q^k-1} \{\alpha V_j\mid \alpha\in\mathbb{F}_{q^n}^*\}, \,\,\mathcal{C}_3=\bigcup\limits_{l=1}^{q^k-1} \{\alpha W_l\mid \alpha\in\mathbb{F}_{q^n}^*\},\,\, \mathcal{C}_4= \{\alpha U\mid \alpha\in\mathbb{F}_{q^n}^*\}.
$$
 Then the code $\mathcal{C}=\mathcal{C}_1\cup\mathcal{C}_2\cup\mathcal{C}_3\cup\mathcal{C}_4$
is a cyclic constant dimension subspace code with cardinality $q^{2k}\frac{q^n-1}{q-1}$ and minimum distance $2k-2$.

\begin{proof}
By Proposition \ref{prop2.5}, we only need to prove that any two distinct subspaces $X$ and $Y$ in the set of $\{ U_i, V_j, W_l, U\}$ satisfy the condition $\dim(X\cap \alpha Y)\leq 1$, for any $\alpha\in \mathbb{F}_{q^n}^*.$  And the theorem can be proved similarly to that in Theorem \ref{thm3.4}.
\end{proof}
\end{thm}

\begin{exam}\label{exam3.8}
Let $n=2\times5\times7=70$, $q=2^4$. The biggest divisor $k$ satisfies that $n/k>2$ is equal to $2\times 7$. Thus $n/k=5>4,$ and $e=\lceil 5/4\rceil -1=1$. Theorem 3.6 in \cite{NXG} can be used to construct a cyclic subspace code with size
$$leq^k\frac{q^n-1}{q-1}=l\times2^{56}\frac{2^{280}-1}{2^4-1}\leq 7\times 2^{57}\frac{2^{280}-1}{2^4-1}.$$
By Theorem \ref{thm3.7} in this paper, we can construct cyclic subspace code with size
$$q^{2k}\frac{q^n-1}{q-1}=2^{112}\frac{2^{280}-1}{2^4-1}\gg 7\times 2^{57}\frac{2^{280}-1}{2^4-1}.$$
\end{exam}

For some positive integers $k, n=r k$ with $r>4$ and a prime power $q$, there is a code constructed in \cite{ZT} with minimum distance $2k$ and size $q^n-1$. In the following Theorem \ref{thm3.11},
we construct a cyclic constant dimension subspace code with minimum distance $2k$ and size $eq^k \frac{q^n-1}{q-1}$.

 Recall that for any two subspaces $U$ and $V$ of $\mathbb{F}_{q^n}$, $U\oplus  V$ denotes the direct sum of $U$ and $V$, and $UV$ is the subspace which is the span of the set of the product of any two elements  $u\in U$ and $v\in V$. We have the following lemma.

\begin{lem}\label{lem3.9}
Let $U, V$ be two distinct subspaces of $\mathbb{F}_{q^n}$ such that $U\cap V=\{0\}$ and $\dim(U\cap \alpha V)\leq 1$, for any $\alpha\in\mathbb{F}_{q^n}^*$. If
\begin{center}
$\mathbb{F}_q+(U+V)+UV=\mathbb{F}_q\oplus(U+V)\oplus UV,$
\end{center}
 then we have
$\dim((\mathbb{F}_q+U)\cap\alpha(\mathbb{F}_q+V))\leq 1$, for any $\alpha\in \mathbb{F}_{q^n}^*.$

\begin{proof}
For any nonzero elements $a+u, a'+u'\in \mathbb{F}_q+U, b+v, b'+v'\in\mathbb{F}_q+V$, assume that $(a+u)(b+v)=(a'+u')(b'+v')$. By hypothesis, we have
\begin{equation}\label{eq3.4}
ab=a'b',
\end{equation}\label{eq3.5}
\begin{equation}
av+bu=a'v'+b'u',
\end{equation}
\begin{equation}\label{eq3.6}
uv=u'v'.
\end{equation}

{\bf Case 1}. $ab\neq 0$ and $uv\neq 0$.
By hypothesis and Proposition 2.2, we have $\frac{u}{u'}=\frac{v'}{v}=\delta\in\mathbb{F}_{q^n}^*$. Assume that $\frac{a}{a'}=\frac{b'}{b}=\lambda\in\mathbb{F}_q^*$. Then Equation (\ref{eq3.5}) can be transformed to $\lambda a'v+\delta b u'=\delta a'v+\lambda b u'.$

If $\lambda=\delta, $ then we have $\frac{a+u}{a'+u'}=\lambda\in\mathbb{F}_q^*$.

If $\lambda\neq\delta, $ then we have $(\lambda-\delta) a'v=(\lambda-\delta) b u'.$ This means that $\frac{u'}{v}=\frac{a'}{b}=\epsilon\in\mathbb{F}_q^*$. This contradicts the condition that $U\cap V=\{0\}$.

{\bf Case 2}. $ab=0, uv\neq 0.$
If $a=0, b=0$, then we can obtain $\frac{u}{u'}=\frac{v'}{v}=\lambda\in\mathbb{F}_q^*$ by hypothesis.

If $a=0, b\neq 0$, we have $a'=0$ or $b'=0$. Assume that $a'=0,$ thus, $\frac{b}{b'}=\frac{u'}{u}=\frac{b+v}{b'+v'}=\lambda\in\mathbb{F}_q^*.$ Assume that $b'=0,$ thus, $\frac{u}{v'}=\frac{a'}{b}=\epsilon\in\mathbb{F}_q^*.$
 This is a contradiction to the condition that $U\cap V=0$.

{\bf Case 3}. $ab\neq 0, uv= 0.$
If $u=0, v=0$, it is easy to obtain that $u'=0, v'=0$. Then we have $\frac{\mathcal{u}}{\mathcal{u}'}=\frac{a}{a'}=\frac{b'}{b}=\lambda\in\mathbb{F}_q^*$ by hypothesis.

If $u=0, v\neq 0$, we have $u'=0$ or $v'=0$. We assume that $u'=0$, then we have $\frac{a}{a'}=\frac{v'}{v}=\frac{\mathcal{u}}{\mathcal{u}'}=\lambda\in\mathbb{F}_q^*.$ We assume that $v'=0$, then we have $\frac{a}{b'}=\frac{a'}{b}=\frac{u'}{v}=\frac{\mathcal{u}'}{\mathcal{v}}=\lambda\in\mathbb{F}_q^*.$ This is a contradiction.

{\bf Case 4}. $ab= 0, uv= 0.$
This means that $a=0, v=0$ or $b=0, u=0$. Assume that $a=0, v=0$, it is easy to obtain that $a'=0, v'=0$ or $b'=0, u'=0.$ If $a'=0, v'=0$, then we have $\frac{\mathcal{u}'}{\mathcal{u}}=\frac{b}{b'}=\frac{u'}{u}=\lambda\in\mathbb{F}_q^*.$ If $b'=0, u'=0,$  then we have $\frac{b}{a'}=\frac{v'}{u}$, this is a contradiction.
The case when $b=0, u=0$ is similarly.
\end{proof}
\end{lem}

In the following lemma, we prove that the sum of $\mathbb{F}_q$ and a Sidon subspace $U$ is again a Sidon space by using Proposition \ref{prop2.6}.

\begin{lem}\label{3.10}
Let $k\geq 2$ be a positive integer and $n=rk, r>4\in\mathbb{Z}^{+},$ and let $e=\lceil r/4 \rceil -1$. Let
\begin{center}
$U_i=\{a+u\gamma+(u^q+\delta_i u)\gamma^{l_i}\mid a\in\mathbb{F}_q, u\in\mathbb{F}_{q^k}\}, i=1,\cdots, e q^{k}$, \end{center}
where $\delta_i\in\mathbb{F}_{q^k}$ and $2\leq l_i \leq e.$ Then $U_i, i=1,\cdots, e q^{k}$, are all Sidon spaces.
\begin{proof}
Write $U_i$ as $\mathbb{F}_q+U'_i$, where $U'=\{u\gamma+(u^q+\delta_i u)\gamma^{l_i}\mid u\in\mathbb{F}_{q^k}\}.$ It is easy to verify that $U'_i$ is a Sidon space. By Proposition \ref{prop2.6}, we just need to prove that $\mathbb{F}_q+U'_i+{U'_i}^2=\mathbb{F}_q \oplus U'_i \oplus {U'_i}^2$. Let
\begin{center}
$\beta_1\in\mathbb{F}_q,$
$\beta_2=u\gamma+(u^q+\delta_i u)\gamma^{l_i}\in U'_i,$
$\beta_3=\sum\limits_{j=1}^{s}\lambda_j u_j v_j\gamma^2+[u_j(v_j^q+\delta_i v_j)+v_j(u_j^q+\delta_i u_j)]\gamma^{l_i+1}+(u_j^q+\delta_i u_j)(v_j^q+\delta_i v_j)\gamma^{2 l_i}\in {U'_i}^2,$
\end{center}
where $s$ is a positive number, and $\beta_1+\beta_2+\beta_3=0$. Since $1,\gamma,\cdots,\gamma^{r-1}$ are linear independent over  $\mathbb{F}_{q^k}$, then $\beta_1+\beta_2+\beta_3=0$ leads to $\beta_1=0$ and $u=0$, which means that $u^q+\delta_i u=0$, and hence $\beta_2=0$. This implies that $\mathbb{F}_q+U'_i+{U'_i}^2=\mathbb{F}_q \oplus U'_i \oplus {U'_i}^2$.
\end{proof}
\end{lem}

\begin{thm}\label{thm3.11}
Let $k\geq 2$ be a positive integer and $n=rk, r>4\in\mathbb{Z}^{+},$ and let $e=\lceil r/4 \rceil -1$. Let
\begin{center}
$U_i=\{a+u\gamma+(u^q+\delta_i u)\gamma^{l_i}\mid a\in\mathbb{F}_q, u\in\mathbb{F}_{q^k}\}, i=1,\cdots, e q^{k}$, \end{center}
where $\delta_i\in\mathbb{F}_{q^k}$ and $2\leq l_i \leq e.$ Then the code
 \begin{center}
 $\mathcal{C}=\bigcup\limits_{i=1}^{e q^{k}} \mathcal{C}_i$,
  \end{center}
  where $\mathcal{C}_i=\{\alpha U_i\mid \alpha\in\mathbb{F}_{q^n}^*\}$, is a cyclic constant dimension subspace code with cardinality $e q^k \frac{q^n-1}{q-1}$ and minimum distance $2k$.
\begin{proof}
We only need to prove that $\dim(U_i\cap \alpha U_j)\leq 1$, for any $\alpha\in\mathbb{F}_{q^n}^*, 1\leq i\neq j\leq e q^{k}.$ Write $U_i$ and $U_j$ as $\mathbb{F}_q+U'_i$ and $\mathbb{F}_q+U'_j$, where $$U'_i=\{u\gamma+(u^q+\delta_i u)\gamma^{l_i}\mid u\in\mathbb{F}_{q^k}\}, U'_j=\{v\gamma+(v^q+\delta_j v)\gamma^{l_j}\mid  v\in\mathbb{F}_{q^k}\}.$$

If $l_i\neq l_j$, then we know that $U'_i\cap U'_j=\{0\}$.

If $l_i=l_j$, since $U_i\neq U_j$, we have $\delta_i\neq \delta_j$. Assume that there exists a nonzero $\mathcal{w}$ in $U'_i\cap U'_j$ and $\mathcal{w}=u\gamma+(u^q+\delta_i u)\gamma^{l_i}=v\gamma+(v^q+\delta_j v)\gamma^{l_i}.$ It follows that
$$u=v, u^q+\delta_i u=v^q+\delta_j v,$$
which means that $\delta_i=\delta_j$. This is a contradiction. We obtain that $U'_i\cap U'_j=0$. By Lemma \ref{lem3.9}, we conclude that $\dim(U_i\cap \alpha U_j)\leq 1$, for any $\alpha\in\mathbb{F}_{q^n}^*, 1\leq i\neq j\leq e q^{k}.$
\end{proof}
\end{thm}

\begin{exam}
 Let $n,k,q$ be the positive integers mentioned  in Example \ref{exam3.8}.  The cyclic constant dimension subspace code constructed by Theorem 3.14 in \cite{ZT} has minimum distance $2k=28$ and size $q^n-1=2^{280}-1$. By Theorem \ref{thm3.11} in this paper, we can construct cyclic constant dimension subspace code with minimum distance $2k=28$ and size $e q^k\frac{q^n-1}{q-1}=2^{56}\frac{2^{280}-1}{2^4-1}\gg 2^{280}-1$.
\end{exam}

\section{Concluding Remarks}

In this paper, we presented several new constructions of large cyclic constant dimension subspace codes by constructing new Sidon spaces. It is shown in Theorem \ref{thm3.4} that we give a construction of code in $\mathcal{G}_q (n, k)$ with size $e q^k (q^n-1)+\frac{q^n-1}{q^k-1}$ when $n/k>2$. In Theorem \ref{thm3.7}, we construct a code in $\mathcal{G}_q (n, k)$ with minimum distance $2k-2$ and size $q^{2k}\frac{q^n-1}{q-1}$ which is much larger than known codes when $n/k>4$. Furthermore, Theorem \ref{thm3.11} shows that we can construct large codes in $\mathcal{G}_q (n, k)$ with increased minimum distance $2k$, as the case $n/k>4$. Our works extend the results in \cite{FW, NXG, ZT}, and cyclic constant dimension subspace codes in this paper have larger sizes than known codes in the previous literature.


\begin{thebibliography}{1}

\bibitem{ACLY} R. Ahlswede, N. Cai, S. Li, R. Yeung, \emph{Network information flow}, IEEE Trans. Inform. Theory,
46(4), 1204-1216 (2000).

\bibitem{BSZ} C. Bachoc, O. Serra, G. Zemor, \emph{An analogue of Vosper's theorem for extension fields},
Math. Proc. Cambridge Philos. Soc., 163(3), 423-452 (2017).

\bibitem{BEGR} E. Ben-Sasson, T. Etzion, A. Gabizon, N. Raviv, \emph{Subspace polynomials and cyclic subspace codes},
IEEE Trans. Inform. Theory, 62(3), 1157-1165 (2016).

\bibitem{BKR} E. Ben-Sasson, S. Kopparty, J. Radhakrishnan, \emph{Subspace polynomials and limits to list decoding of Reed-Solomon codes},
IEEE Trans. Inform. Theory, 56(1), 113-120 (2010).

\bibitem{CL} B. Chen, H. Liu, \emph{Constructions of cyclic constant dimension codes}, Des. Codes Cryptogr.,
86(6), 1267-1279 (2017).

\bibitem{EV} T. Etzion, A. Vardy, \emph{Error-correcting codes in projective space},
IEEE Trans. Inform. Theory, 57(2), 1165-1173 (2011).

\bibitem{FW} T. Feng, Y. Wang, \emph{New constructions of large cyclic subspace codes and Sidon spaces},
Discrete Math., 344(4), 112273 (2021).

\bibitem{GL} H. Gluesing-Luerssen, H. Lehmann, \emph{Distance distributions of cyclic orbit codes},
Des. Codes Cryptogr., 89(3), 447-470 (2021).

\bibitem{GMT} H. Gluesing-Luerssen, K. Morrison, C. Troha, \emph{Cyclic orbit codes and stabilizer subfields},
Adv. Math. Commun., 9(2), 177-197 (2015).

\bibitem{KKS} A. Kohnert, S. Kurz, \emph{Constructing of large constant dimension codes with a prescribed minimum distance},
Math. Methods Comput. Sci., 5393, 31-42 (2008).

\bibitem{KK} R. K\"otter, F. R. Kschischang, \emph{Coding for errors and erasures in random network coding},
IEEE Trans. Inform. Theory, 54(8), 3579-3591  (2008).

\bibitem{LC} H. Lao, H. Chen, {\em  New constant dimension subspace codes from multilevel linkage construction}, Adv. Math. Commun., doi:10.3934/amc.2022039 (2022).

\bibitem{LL} Y. Li, H. Liu, \emph{Cyclic constant dimension subspace codes via the sum of Sidon
spaces}, Des. Codes Cryptogr.,  doi:10.1007/s10623-022-01146-9  (2022).


\bibitem{LN} R. Lidl, H. Niederreiter, Finite Fields, Cambridge University Press, Cambridge, (1997).

\bibitem {NXG} M. Niu, J. Xiao, Y. Gao, {\em New constructions of large cyclic subspace codes via Sidon spaces}, Adv. Math. Commun., doi:10.3934/amc.2022074 (2022).

\bibitem{NYW} Y. Niu, Q. Yue, Y. Wu, \emph{Several kinds of large cyclic subspace codes via Sidon spaces},
Discrete Math., 343(5), 111788 (2020).

\bibitem{OO} K. Otal, F. {\"O}zbudak, \emph{Cyclic subspace codes via subspace polynomials}, Des. Codes Cryptogr.,
85(2), 191-204 (2017).

\bibitem{RT} J. Rosenthal, A. L. Trautmann, \emph{A complete characterization of irreducible cyclic orbit codes and their Plücker embedding},
Des. Codes Cryptogr., 66(1-3), 275-289 (2013).

\bibitem{RRT} R. Roth, N. Raviv, I. Tamo, \emph{Construction of Sidon spaces with applications to coding},
IEEE Trans. Inform. Theory, 64(6), 4412-4422 (2017).


\bibitem{T} A. L. Trautmann, \emph{Isometry and automorphisms of constant dimension codes},
Adv. Math. Commun., 7(2), 147-160 (2013).

\bibitem{TMBR} A. L. Trautmann, F. Manganiello, M. Braun, J. Rosenthal, \emph{Cyclic orbit codes},
IEEE Trans. Inform. Theory, 59(11), 7386-7404 (2013).

\bibitem{XF} S. Xia, F. Fu, \emph{Johnson type bounds on constant dimension codes},
Des. Codes Cryptogr., 50(2), 163-172 (2009).

\bibitem{ZT} H. Zhang, C. Tang, {\em Constructions of large cyclic constant dimension codes via Sidon spaces}. Des. Codes Cryptogr., doi:10.1007/s10623-022-01095-3  (2022).

\end{thebibliography}
\end{document}